\documentclass[a4paper]{panl}
\usepackage{cite}
\usepackage{wrapfig}
\usepackage{graphicx}
\usepackage{amssymb}
\usepackage{amsfonts}
\usepackage{amsmath}
\usepackage{longtable}
\usepackage{rotating}
\usepackage{lscape}
\usepackage{epsfig}
\usepackage{multirow}
\usepackage{caption}
\usepackage{subcaption}
\usepackage{float}
\usepackage{slashbox}

\def\tr{\mathrm{tr}}

\originalTeX
\begin{document}

\title{Critical exponents of model with matrix order parameter from resummation of six-loop results for anomalous dimensions}
\maketitle
\authors{N.M.\,Lebedev$^{\ a}$\footnote{E-mail: lebedev@theor.jinr.ru}}
\setcounter{footnote}{0}
\from{$^{a}$\,Bogoliubov Laboratory of Theoretical Physics,
Joint Institute for Nuclear Research,
141980 Dubna, Moscow region, Russia}

\begin{abstract}

In this contribution an application of two techniques for resummation of asymptotic series namely Borel-Pade technique and Borel-Leroy technique with conformal mapping to the case of a model with multiple coupling constants will be discussed and the results of application of this methods to the $O(n)$-symmetric $\phi^{4}$ model with an antisymmetric tensor order parameter will be presented.
\end{abstract}
\vspace*{6pt}


\label{sec:intro}
\section*{Introduction}

We study critical behaviour of the $O(n)$-symmetric $\phi^{4}$ model with an antisymmetric tensor order parameter $\phi_{ik}=-\phi_{ki}$; $i$, $k=1$, \dots, $n$. The action of the model:
\begin{equation}
\label{action}
S(\phi)=\frac{1}{2}\,\tr\left(\phi\left(-\partial^{2}+m^{2}_{0}\right)\phi\right) - \frac{g_{10}}{4!} \left(\tr\left(\phi^{2}\right)\right)^{2} - \frac{g_{20}}{4!}\, \tr \left(\phi^{4}\right).
\end{equation}
includes two independent $O(n)$ invariant quartic structures and consequently two independent coupling constants. Previously, this model was studied in the framework of minimal subtraction (MS) scheme with  $\varepsilon$-expansion and renormalization group approach in the space of a fixed dimension with pseudo-$\varepsilon$-expansion up to four-loop order \cite{AKL1,KL1,KL2}. It was shown that requirement of convergence of a functional integral with action (\ref{action})  imposes restriction on the values the couplings can take:
\begin{eqnarray}
\label{ineq}
&\text{even}& \qquad 2 g_{10} + g_{20} >0, \quad n g_{10} + g_{20} >0, \\
&\text{odd} &\qquad 2 g_{10} + g_{20} >0, \quad (n - 1) g_{10} + g_{20} >0, \nonumber
\end{eqnarray}
and also high order asymptotic (HOA) for coefficients of the perturbation series and corresponding $\varepsilon$-expansion had been found:
\begin{equation}
    \label{hoag}
    \beta^{(N)}_{i}(g_{1},g_{2})= \text{Const} \cdot N!N^{b}(-a(g_{1},g_{2}))^{N}\left(1+O\Big(\frac{1}{N}\Big)\right)
\end{equation}
\begin{equation}
    \label{hoae}
    g^{(N)}_{1,2*} = \text{Const} \cdot N!N^{b+1}(-a(g^{(1)}_{1*},g^{(1)}_{2*}))^{N}\left(1+O\Big(\frac{1}{N}\Big)\right).
\end{equation}
Here $a(g_{1},g_{2})=\max_{k}\ [a_{k}(g_{1},g_{2})]=\max_{k}\,((2kg_{1} + g_{2})/4k)$; $k=1$, \dots, $n/2$; $g^{(1,2)}_{1*}$ is a one loop contribution to a coordinates of the fixed points and $a_{k}$ correspond to different instanton solutions.

It is also know that in cases $n=2,3$ the model is reduced to scalar and $O(3)$-vector models, and for $n>4$ there is no IR stable fixed points within perturbation theory. In case $n=4$ there are 3 fixed points: A which is a saddle point at all orders and points B  and C which are saddle point and IR stable point at the 4-loop level. The coordinates of the latter point are known to be the subject of the relation $g^{*}_{1} = -0.75 g^{*}_{2}$ at all orders.

Recently the $\varepsilon$-expansions were extended up to six order \cite{PB} which allows us refine our understanding of the critical properties of the model and provides a great sandbox to study stability of certain resummation techniques based on Borel-Leroy transformation in the case of multi charge model.

\label{sec:borel-leroy}
\section*{Borel-Leroy transformation}

For some quantity $f(z)$ Borel-Leroy transform is defined as follows:
\begin{equation}
    f(z) = \sum_{N \ge 0} f_{N} z^{N}; \quad  \Rightarrow \quad    B(t) = \sum_{N \ge 0} \frac{f_{N}}{\Gamma(N+b_{0}+1)}\ t^{N} = \sum_{N \ge 0} B_{N} t^{N}.
\end{equation}
If the original series was asymptotic with exponentially growing coefficients then series for Borel image will be convergent in a circle of the radius $1/a$. 
\begin{equation}
    f_{N} \simeq \text{Const} \cdot N!N^{b}(-a)^{N} \quad \Rightarrow \quad B_{N} \simeq \text{Const} \cdot N^{b - b_{0}}(-a)^{N}.
\end{equation}
So that in order to perform inverse transform and get resumed quantity:
\begin{equation}
    f^{\mathrm{res}}(z) = \int^{\infty}_{0} dt\ t^{b_o} e^{-t} B(tz)
\end{equation}
one should construct an analytical continuation for $B(t)$ outside of the convergence radius (for a more detailed discussion of Borel-Leroy-based techniques see \cite{K} and references therein).

\label{sec:Conformal-borel}
\section*{Conformal-Borel}

One possible way to perform inverse transform is to map integration contour inside the convergence radius of $B(t)$:
\begin{equation}
u(t) = \frac{\sqrt{1+at}-1}{\sqrt{1+at}+1} \quad \Leftrightarrow \quad t(u) = \frac{4u}{a(u-1)^{2}},
\label{mapping}
\end{equation}
where $a$ is the parameter of HOA (\ref{hoag}), (\ref{hoae}) then re-expand in terms of new variable and perform inverse transform. The choice of this mapping function together with setting free parameter $b_{0} = 3/2$ guaranties that resumed series will have correct HOA. Values of critical exponents obtained in such a way presented in Tables \ref{tabconfpa}--\ref{confpc}. 

\begin{table}[H]
\caption{\label{tabconfpa}Values of critical exponents at different number of loops taken into account for the fixed point A.}
\centering
\begin{tabular}{|c||c|c|c|c|c|c|}
\hline
\multirow{2}{*}{quantity} & \multicolumn{3}{c|}{$d=2$} & \multicolumn{3}{c|}{$d=3$} \\
\cline{2-7}
& 4 loop & 5 loops & 6 loops & 4 loops & 5 loops & 6 loops\\
\hline \hline
$g_{1}^{*}$ & 1.10 & 1.24 & 1.34 & 0.542 & 0.571 & 0.586 \\
\hline
$g_{2}^{*}$ & 0 & 0 & 0 & 0 & 0 & 0 \\
\hline
$\omega_1$ & $-0.487$ & $-0.583$ & $-0.673$ & $-0.226$ & $-0.246$ & $-0.260$\\
\hline
$\omega_2$ & 1.35 & 1.39 & 1.36 & 0.781 &  0.791 & 0.786\\
\hline
$\eta$ & 0.0557 & 0.0820 & 0.106 & 0.0192 & 0.0245 & 0.0279\\
\hline
\end{tabular}

\bigskip

\caption{\label{confpb}Values of critical exponents at different number of loops taken into account for the fixed point B.}
\centering
\begin{tabular}{|c||c|c|c|c|c|c|}
\hline

\multirow{2}{*}{quantity} & \multicolumn{3}{c|}{$d=2$} & \multicolumn{3}{c|}{$d=3$} \\
\cline{2-7}
& 4 loop & 5 loops & 6 loops & 4 loops & 5 loops & 6 loops\\
\hline \hline
$g_{1}^{*}$ & 2.40 & 2.89 & 3.21 & 1.14 & 1.26 & 1.32 \\
\hline
$g_{2}^{*}$ & $-3.95$ & $-5.21$ & $-6.32$ & $-1.75$ & $-2.04$ & $-2.24$ \\
\hline
$\omega_1$ & $-0.198$ & $-0.413$ & $-0.683$ & $-0.0547$ & $-0.105$ & $-0.153$\\
\hline
$\omega_2$ & 1.28 & 1.36 & 1.43 & 0.755 & 0.774 & 0.787\\
\hline
$\eta$ & 0.0126 & 0.0267 & 0.0440 & 0.00407 & 0.00721 & 0.0101\\
\hline
\end{tabular}

\bigskip

\caption{\label{confpc}Values of critical exponents at different number of loops taken into account for the fixed point C.}
\centering
\begin{tabular}{|c||c|c|c|c|c|c|}
\hline

\multirow{2}{*}{quantity} & \multicolumn{3}{c|}{$d=2$} & \multicolumn{3}{c|}{$d=3$} \\
\cline{2-7}
& 4 loop & 5 loops & 6 loops & 4 loops & 5 loops & 6 loops\\
\hline \hline
$g_{1}^{*}$ & 2.02 & 2.32 & 2.54 & 1.03 & 1.10 & 1.14 \\
\hline
$g_{2}^{*}$ & $-2.69$ & $-3.10$ & $-3.38$ & $-1.37$ & $-1.47$ & $-1.52$ \\
\hline
$\omega_1$ & 0.245 & 0.377 & 0.500 & 0.0774 & 0.109 & 0.131\\
\hline
$\omega_2$ & 1.30 & 1.38 & 1.41 & 0.762 & 0.781 & 0.787\\
\hline
$\eta$ & 0.0477 & 0.0744 & 0.101 & 0.0177 & 0.0238 & 0.0284\\
\hline
\end{tabular}
\end{table}

\label{sec:pade-borel}
\section*{Pade-Borel}

More simple way to make analytical continuation of Borel image is to construct Pade approximants of it in such a way that the initial terms of the series
expansion of $B(t)$ are reproduced.
\begin{equation}
    f^{\mathrm{res}}_{[N, M]}(z) = \int^{\infty}_{0}\!\! dt\ e^{-t} t^{b_o} B_{[N,M]}(zt); \quad B_{[N, M]}(t) = \frac{P_{N}(t)}{P_{M}(t)} = \frac{\sum_{i =0}^{i = N} \alpha_{i} t^{i}}{\sum_{j =0}^{j = M} \beta_{j}t^{j}}\,.
\end{equation}
Values of critical exponents corresponding to the IR stable fixed point at $d=3$ obtained in such a way presented in Tables~\ref{tabeps}--\ref{tabeps4}. 

\begin{table}[H]
\caption{\label{tabeps}Coordinate $g^{*}_{1}$.}
\centering
\begin{tabular}{|c||c c c c c c }
\hline
\backslashbox{N}{M} & 0 & 1 & 2 & 3 & 4 & 5  \\
\hline 
1 & 0.818 & 2.93 & 0.772 & 1.44 & 0.471 & 0.0260 \\
\cline{1-1}
2 & 1.28 & 1.12 & 1.23 & 1.12 & 1.32& \\
\cline{1-1}
3 & 1.03 & 1.19 & 1.17 & 1.18 &  & \\
\cline{1-1}
4 & 1.51 & 1.16 & 1.18 &  &  & \\
\cline{1-1}
5 & 0.222 & 1.20 &  &  &  &\\
\cline{1-1}
6 & 4.48 &  &  &  &  &  \\
\cline{1-1}
\end{tabular}
\end{table}
\begin{table}[H]
\caption{\label{tabeps2}Eigenvalue $\omega_{1}$.}
\centering
\begin{tabular}{|c||c c c c c c }
\hline
\backslashbox{N}{M} & 0 & 1 & 2 & 3 & 4 & 5  \\
\hline 
1 & $-0.0909$ & $-0.0217$ & $-0.00783$ & $-0.00313$ &  $-0.00143$ & $-0.00072$ \\
\cline{1-1}
2 & 0.221 & 0.0912 & 0.0337 & 0.0530 & $-0.0377$ & \\
\cline{1-1}
3 & $-0.00926$ & 0.153 & 0.156 & 0.210 &  & \\
\cline{1-1}
4 & 0.578 & 0.156 & 0.153 &  &  & \\
\cline{1-1}
5 & $-1.00$ & 0.199 &  &  &  &\\
\cline{1-1}
6 & 4.28 &  &  &  &  &  \\
\cline{1-1}
\end{tabular}
\end{table}
\begin{table}[H]
\caption{\label{tabeps3}Eigenvalue $\omega_{2}$.}
\centering
\begin{tabular}{|c||c c c c c c }
\hline
\backslashbox{N}{M} & 0 & 1 & 2 & 3 & 4 & 5  \\
\hline 
1 & 1.00 & 0.645 & 1.18 & 0.434 & $-0.129$ &  0.201 \\
\cline{1-1}
2 & 0.430 & 0.817 & 0.776 & 0.818 & 0.755 & \\
\cline{1-1}
3 & 1.71 & 0.769 & 0.797 & 0.791 &  & \\
\cline{1-1}
4 & $-2.07$ & 0.831 & 0.790 &  &  & \\
\cline{1-1}
5 & 11.1 & 0.708 &  &  &  &\\
\cline{1-1}
6 & $-41.1$ &  &  &  &  &  \\
\cline{1-1}
\end{tabular}
\end{table}
\begin{table}[H]
\caption{\label{tabeps4}Exponent $\eta$.}
\centering
\begin{tabular}{|c||c c c c c }
\hline
\backslashbox{N}{M} & 0 & 1 & 2 & 3 & 4  \\
\hline 
2 & 0.0206 & 0.0802 & 0.0170 & $-0.00749$ & 0.00832 \\
\cline{1-1}
3 & 0.0391 & 0.0339 & 0.0429 & 0.0339 & \\
\cline{1-1}
4 & 0.0316 & 0.0370 & 0.0372 &  & \\
\cline{1-1}
5 & 0.0520 & 0.0372 &  &  &\\
\cline{1-1}
6 &  $-0.00503$ &  &  &  &  \\
\cline{1-1}
\end{tabular}
\end{table}

\label{sec:Proximity}
\section*{Proximity of the ressumed series to the exact results}
If we calculate analytical continuation of $B(t)$ from first $l$ known coefficients we can expand it back in powers of $t$ to find that the expansion not just reproduce first $l$ coefficients but also add some additional sub-series that we are actually summing up.
\begin{equation}
    B_{\mathrm{continued}}(t) = \sum_{N \leq l} B_{N} t^{N} + \sum_{N > l} B^{r}_{N} t^{N}.
\end{equation}
In order to estimate how close this reconstructed sub-series to the unknown exact coefficients one can try to reconstruct last know coefficient $B_{l}$ taking into account less of know contributions, and then estimate proximity to exact answer and convergence rate calculating relative discrepancy from exact value. 
\begin{equation}
    \xi_{l} = \frac{f_{l} - f^{r}_{l}}{f_{l}}\,.
\end{equation}
The estimates of the value of $\xi_{6}$ for $\varepsilon$-expansions fixed point C obtained from conformal mapping are presented in Table~\ref{predictA} and for Pade approximation in Tables~\ref{tabeps5}--\ref{tabeps6}.

\begin{table}[H]
\caption{\label{predictA}$\xi_{6}$ value for $\varepsilon$-expansions at the IR attractive fixed point.}
\centering
\begin{tabular}{|c|c|c|c|}
\hline
 quantity & 3 loops & 4 loops & 5 loops \\
 \hline
 $g^{*}_{1}$ & $-41.9$ & 18.2 & $-3.15$ \\
  \hline
 $\omega_{1}$ & $-9.47$ & 5.93 & $-1.47$  \\
  \hline
 $\omega_{2}$ & 0.996 & 0.424 & 0.0389  \\
  \hline
 $\eta$ & 100 & $-80.7$ & 22.7 \\
 \hline
\end{tabular}
\end{table}

\begin{table}[H]
\caption{\label{tabeps5}$\xi_{6}$ value for $\varepsilon$-expansions of $g^{*}_{1}$ (left) and $\eta$ (right) at the IR attractive fixed point.}
\begin{tabular}{|c||c c c c c }
\hline
\backslashbox{N}{M} & 1 & 2 & 3 & 4  \\
\hline 
1 & 0.973 & 0.889 & 0.998 & 1.97  \\
\cline{1-1}
2 & 0.983 & 0.700 & 0.270 &  \\
\cline{1-1}
3 & 0.500 & 0.113 &  & \\
\cline{1-1}
4 & 0.136 &  &  & \\
\cline{1-1}
\end{tabular}
\quad
\begin{tabular}{|c||c c c c }
\hline
\backslashbox{N}{M} & 1 & 2 & 3  \\
\hline 
2 & 1.38 & 0.525 & 2.74 \\
\cline{1-1}
3 & 0.973 & 0.469 &  \\
\cline{1-1}
4 & $-0.0563$ &  & \\
\cline{1-1}
\multicolumn{1}{c}{} & & &
\end{tabular}
\end{table}

\begin{table}[H]
\caption{\label{tabeps6}$\xi_{6}$ value for $\varepsilon$-expansions of $\omega_{2}$ at the IR attractive fixed point.}
\centering
\begin{tabular}{|c||c c c c c }
\hline
\backslashbox{N}{M} & 1 & 2 & 3 & 4  \\
\hline 
1 & 0.997 & 0.915 & 0.708 & 0.495 \\
\cline{1-1}
2 & 0.550 & 0.194 & 0.0432 &  \\
\cline{1-1}
3 & 0.212 & 0.0197 &  & \\
\cline{1-1}
4 & 0.0550 &  &  & \\
\cline{1-1}
\end{tabular}
\end{table}

\label{sec:large eps}
\section*{Large $\varepsilon$ behaviour}
\begin{figure}[H]
\centering
\begin{subfigure}[t]{65mm}
  \includegraphics[width=65mm]{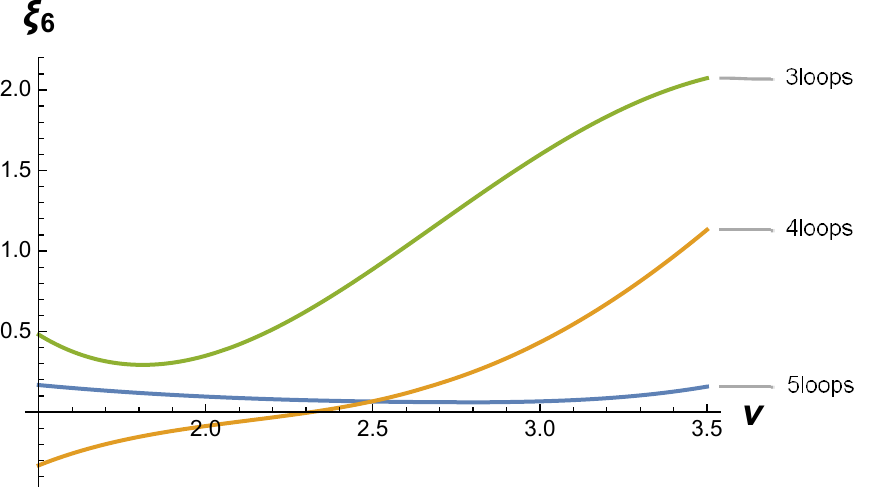}
  \caption{}
\end{subfigure}
\hfill
\begin{subfigure}[t]{65mm}
  \includegraphics[width=65mm]{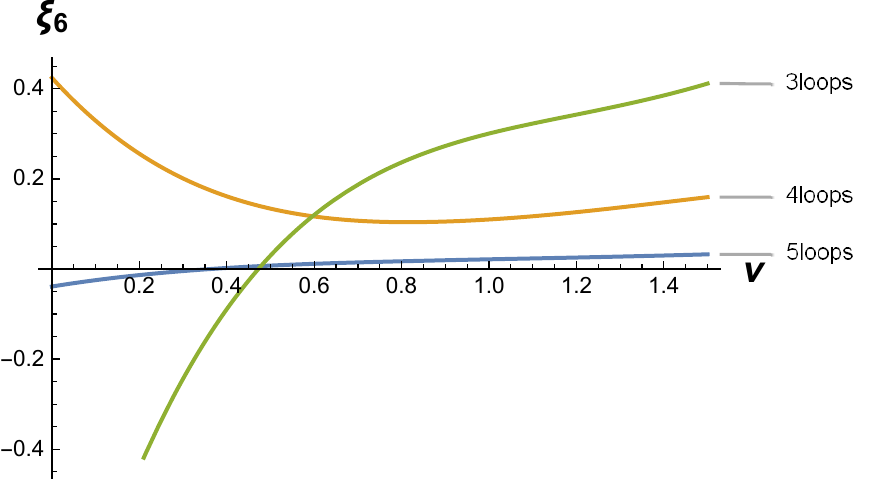}
  \caption{}
\end{subfigure}
\begin{subfigure}[t]{65mm}
  \includegraphics[width=65mm]{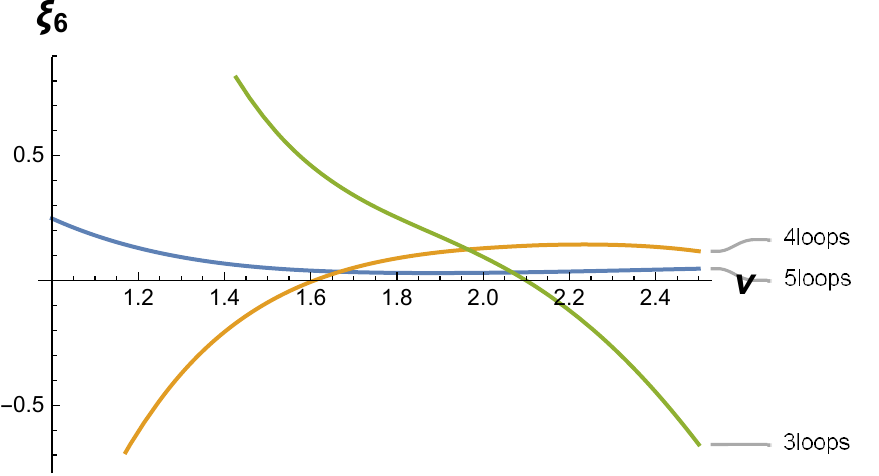}
  \caption{}
\end{subfigure}
\hfill
\begin{subfigure}[t]{65mm}
  \includegraphics[width=65mm]{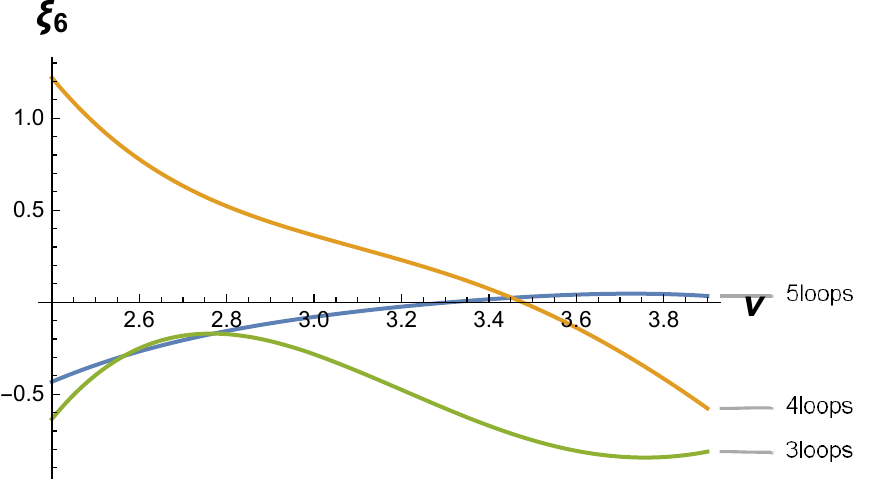}
  \caption{}
\end{subfigure}
\caption{\label{fig1} Dependence of the relative discrepancy $\xi_{6}$ on the value of $\nu$ for $\omega_{1}$ (a), $\omega_{2}$ (b), $g^{*}_{1}$ (c), $\eta$ (d) for IR stable fixed point with the different number of loops taken into account.}
\end{figure}

Since the mapping function (\ref{mapping}) tends to unity at large values of $t$
it is possible to modify conformal analytical continuation in a way that allows one to control not only HOA but also large $z$ behaviour of the resumed series 
\begin{equation}
    \widetilde{B}(u(t)) = \bigg[\frac{t}{u(t)}\bigg]^{\nu} \sum_{N \leq l} B_{N} u(t)^{N}.
\end{equation}
It's shown \cite{K} that if exact series has power asymptotic then setting parameter $\nu$ close to its actual value speeds up rate of order by order convergence. More over in case of scalar $\phi^{4}$ model relative discrepancy $\xi_{l}$ tends to have minimal absolute value at the actual value of $\nu$. However from the Figure~\ref{fig1} it can be seen that in our case there seem to be no universal value of $\nu$, even for different $\varepsilon$-series separately, in which vicinity $\xi_{6}$ minimizes its absolute value.

\label{sec:beta res}
\section*{Ressumation of $\beta$ functions}
Except for resummation of $\varepsilon$-expansions we can also resum directly series of $\beta$-functions in the powers of couplings. For that purpose we should rescale couplings:
\begin{equation}
    \beta(g_{1}, g_{2}) = \sum_{i,j} \beta_{i,j} \ g_{1}^{i} g_{2}^{j} \quad \Rightarrow \quad \beta(z) = \beta(z g_{1}, z g_{2}) = \sum_{i} \beta_{i}(g_{1}, g_{2})z^{i}
\end{equation}
so that $\beta(z)|_{z=1}=\beta(g_{1}, g_{2})$, and then resum the series in powers of $z$ with coefficients depending on couplings.
\begin{equation}
    B(t) = \sum \frac{\beta_{N}(g1, g2)}{\Gamma(N+b_{0}+1)}\ t^{N} = \sum B_{N}(g_{1}, g_{2}) t^{N}.
\end{equation}
At each point of couplings plane the most relevant instanton shell be used so that value of parameter a in (\ref{mapping}) is now depend on the particular coordinate on the plane of invariant couplings.
Ressumed $\beta$-functions are given by:
\begin{equation}
    \beta^{\mathrm{res}}(g_{1}, g_{2}) = \beta^{\mathrm{res}}(z=1) = \int^{\infty}_{0} dt\ e^{-t} t^{b_o} B(t).
\end{equation}
Finaly we solve numerically system of equations for invariant coupling:
\begin{equation}
    s\partial_{s}\bar g(s,g) = \beta^{\mathrm{res}}_{g} (\bar g), \quad
\bar g(1,g) = g \quad s = p/\mu.
\end{equation}

\begin{figure}[H]
\begin{subfigure}[t]{65mm}
  \includegraphics[width=70mm]{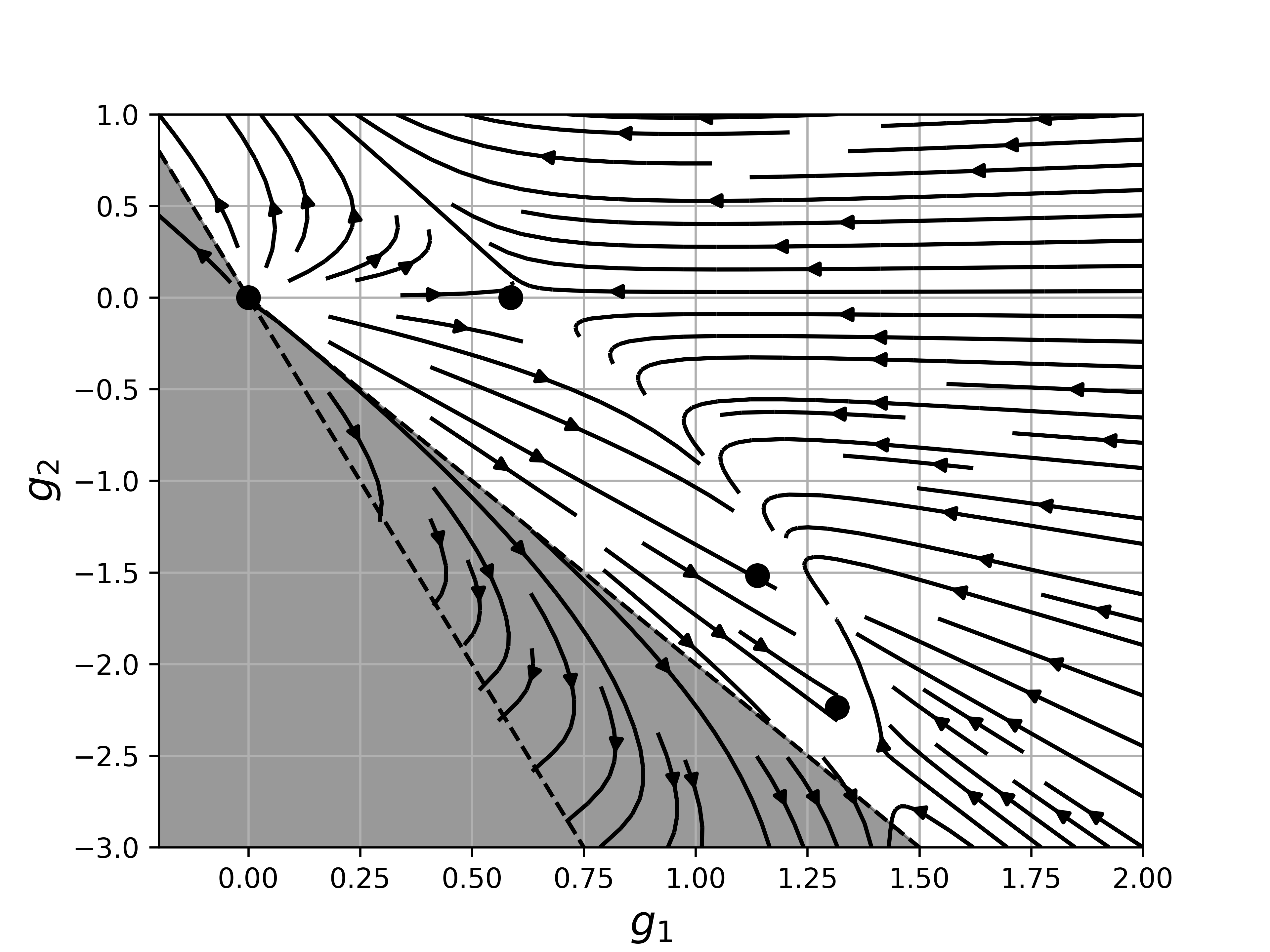}
  \caption{$d=2$}
\end{subfigure}
\hfill
\begin{subfigure}[t]{65mm}
  \includegraphics[width=70mm]{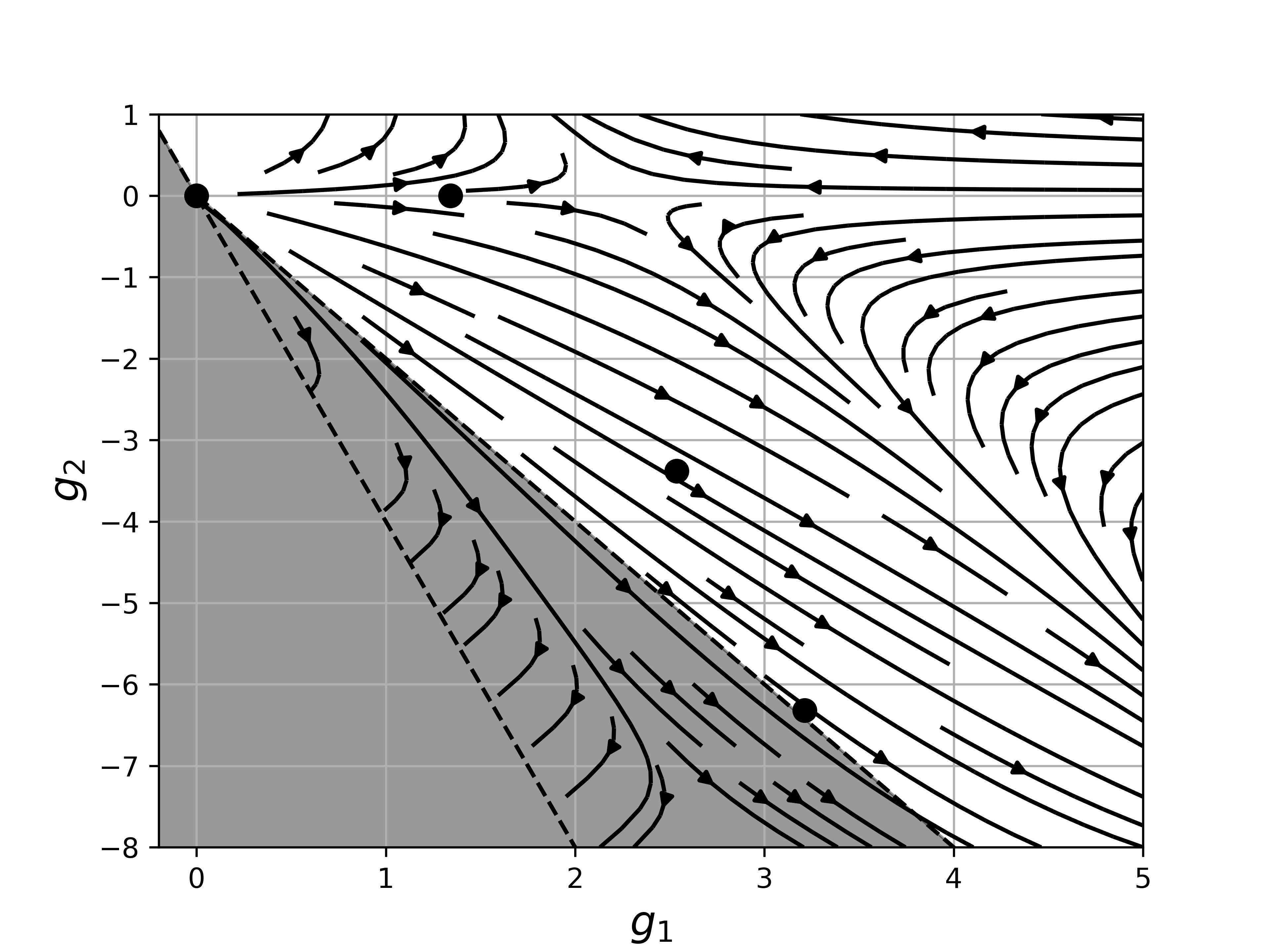}
  \caption{$d=3$}
\end{subfigure}
\caption{Ressumed RG flows in the plane of invariant couplings. The gray area is unphysical region according to (\ref{ineq}) lower dashed line separates the region where instanton solution governing asymptotics of perturbative series formally melts down making resummation procedure meaningless. The black dots mark fixed points positions according to Table~\ref{confpc}.}\label{fig2}
\end{figure}

As can be seen from the Fig.~\ref{fig2}(a) at $d=3$ we have only qualitatively agreement between results of resummation of $\varepsilon$-expansions for fixed points and root positions of $\beta$-functions resumed directly. At the same time Fig.~\ref{fig2}(b) shows that at $d=2$ situation seem to be much worse because resumed $\beta$-functions are lacking of two nontrivial fixed points at all. The later is indicating that for $d=2$ six orders or perturbation theory may be not sufficient to make even consistent qualitative conclusions about asymptotic regimes of the model under consideration based on the resummation techniques.

\label{sec:Conclusions}
\section*{Conclusions}

We have obtained coordinates of the fixed points and established their IR stability properties at 6-loop level resuming corresponding $\varepsilon$-expansions using two different procedures based on the Borel-Leroy transform. For one of the points that appeared to be IR attractive we also calculated anomalous dimension of a pair correlation function. These results are in qualitative agreement with each other but numerically show discrepancy already at the level of a second significant digit. We obtained estimates for how close first unknown coefficients of resumed series are to their exact values. The best predictions are obtained for $\omega_{2}$ eigenvalue which original $\varepsilon$-series are most divergent, while the highest discrepancy is obtained for $\eta$ series which is still showing apparent convergence even at 6-loop level. We have shown that accounting for possible power-like asymptotic of $\varepsilon$-expansions does not help to optimise resummation procedure. And finally we have resumed expansions for $\beta$-functions directly and shown that at $d=3$ they qualitatively agree with results of $\varepsilon$-expansions resummation while at $d=2$ there is no even qualitatively agreement. The precise reason of such a discrepancy could be a subject of a separate study.

\label{sec:Acknowledgments}
\section*{Acknowledgments}
Author is grateful to A.F. Pikelner and G.A. Kalagov for useful discussions and comments and to FFK2021 Organizing Committee for support and hospitality. The reported study was funded by the Russian Foundation for Basic Research, project number 20-32-70139.


\end{document}